\documentclass[conference]{IEEEtran}

\IEEEoverridecommandlockouts
\usepackage{cite}
\usepackage{amsmath,amssymb,amsfonts}
\usepackage{algorithmic}
\usepackage{graphicx}
\usepackage{textcomp}
\usepackage{xcolor}
\usepackage{soul}

\usepackage{dcolumn}
\usepackage{subfigure}
\usepackage{float}
\usepackage{caption}
\usepackage{tabularx}

\usepackage{stfloats}

\def\BibTeX{{\rm B\kern-.05em{\sc i\kern-.025em b}\kern-.08em
    T\kern-.1667em\lower.7ex\hbox{E}\kern-.125emX}}

\newcommand{\scrap}[1]{\textcolor{black}{}}
\newcommand{\insertred}[1]{{#1}}

\begin{document}

\title{End-to-end learnable EEG channel selection for deep neural networks \insertred{with Gumbel-softmax} \\}
\author{\IEEEauthorblockN{Thomas Strypsteen and Alexander Bertrand}
\IEEEauthorblockA{\textit{KU Leuven, Department of Electrical Engineering (ESAT)} \\
\textit{STADIUS Center for Dynamical Systems, Signal Processing and Data Analytics}\\
Leuven, Belgium \\
\{thomas.strypsteen, alexander.bertrand\}@esat.kuleuven.be}

\thanks{This project has received funding from the European Research Council (ERC) under the European Union’s Horizon 2020 research and innovation
programme (grant agreement No 802895). The authors also acknowledge the financial support of the FWO (Research Foundation Flanders) for project G.0A49.18N, and the Flemish Government under the “Onderzoeksprogramma Artifici\"ele Intelligentie (AI) Vlaanderen” programme.
\newline
T. Strypsteen and A. Bertrand are with KU Leuven, Department of Electrical Engineering (ESAT), STADIUS Center for Dynamical Systems, Signal Processing and Data Analytics and with Leuven.AI - KU Leuven institute for AI, Kasteelpark Arenberg 10, B-3001 Leuven, Belgium (e-mail: thomas.strypsteen@kuleuven.be, alexander.bertrand@kuleuven.be).}
}

\maketitle

\begin{abstract}
\textit{Objective.} To develop an efficient, embedded electroencephalogram (EEG) channel selection approach for deep neural networks, allowing us to match the channel selection to the target model, while avoiding the large computational burdens of wrapper approaches in conjunction with neural networks.
\textit{Approach.} We employ a concrete selector layer to jointly optimize the EEG channel selection and network parameters. This layer uses a Gumbel-softmax trick to build continuous relaxations of the discrete parameters involved in the selection process, allowing them be learned in an end-to-end manner with traditional backpropagation. As the selection layer was often observed to include the same channel twice in a certain selection, we propose a regularization function to mitigate this behavior. We validate this method on two different EEG tasks: motor \insertred{execution} and auditory attention decoding. For each task, we compare the performance of the Gumbel-softmax method with a baseline EEG channel selection approach tailored towards this specific task: mutual information and greedy forward selection with the utility metric respectively.
\textit{Main results.} Our experiments show that the proposed framework is generally applicable, while performing at least as well as (and often better than) these state-of-the-art, task-specific approaches. 
\textit{Significance.} The proposed method offers an efficient, task- and model-independent approach to jointly learn the optimal EEG channels along with the neural network weights.
\end{abstract}


\section{Introduction}

Electroencephalography (EEG) is a widely used neuro-monitoring technique that measures the brain's electrical activity in a noninvasive way. Its applications are numerous, including detection of epileptic seizures \cite{ansari2019neonatal}, monitoring sleeping patterns \cite{de2017complexity}, studying brain disorders after injuries \cite{giacino2014disorders}, providing communication means for motor-impaired patients through brain-computer interfaces (BCIs) \cite{lawhern2018eegnet} and many more. However, acquiring these EEG signals typically involves wearing bulky, heavy EEG caps containing a large amount of electrodes with conductive gel, resulting in an uncomfortable user experience and restricting the monitoring to hospital or lab settings.
These limitations of classical EEG have led to a growing desire for ambulatory EEG, allowing for continuous neuromonitoring in daily life \cite{casson2010wearable}. This shift to mobile applications means the EEG cap is replaced by a number of lightweight, concealable mini-EEG devices, possibly organized in a wireless EEG sensor network (WESN) \cite{bertrand2015distributed}, \cite{narayanan2019analysis}, \cite{tang202034}. Since recording and transmitting all possible channels would incur enormous energy costs, selecting an optimal subset of channels to perform the given task constitutes a crucial step in this wireless setting. Even in more traditional EEG settings, reducing the number of channels offers numerous advantages: it reduces the setup time in clinical settings, helps prevent overfitting effects, decreases the computational load and improves interpretability of the model by removing uninformative channels.
\newline

The last few years, deep learning models have emerged as a popular EEG analysis tool \cite{roy2019deep}. For several applications, it has been shown that replacing the classical signal processing approaches with deep neural networks (DNNs) can substantially improve performance \cite{vandecappelle2019eeg, ansari2019neonatal, lawhern2018eegnet}. In this paper, we focus on the EEG channel selection problem in DNNs. A major problem that arises when performing channel selection - which can be viewed as a grouped feature selection - for neural networks is that many popular feature selection techniques are wrapper approaches. This means that a certain heuristic search is performed on the space of possible feature subsets, training a model on each of these candidate subsets and selecting the one where the model's performance is optimal. However, training a neural network is computationally a lot more demanding than traditional machine learning algorithms, rendering these procedures far too time-consuming for practical use. 
\newline

Embedded approaches, that aim to learn the channel selection jointly with the network weights in an end-to-end manner might offer a more promising avenue in this regard. To deal with the discrete parameters inherent to such a subset selection problem, embedded feature selection techniques can employ the Gumbel-softmax trick \cite{jang2016categorical} to build continuous approximations of these parameters, resulting in the concept of a \textit{concrete selector layer} \cite{abid2019concrete,singh2020fsnet}. This layer can be prepended to any given neural network model to learn the optimal feature selection for the given task. In this work, we investigate the use of these Gumbel-softmax based approaches for the case of EEG channel selection. We demonstrate that straightforwardly applying this approach to EEG channel selection often leads to the selection of duplicate channels since the selection neurons in this layer act independently and are not 'aware' of each others' selection. In high-dimensional data sets as in \cite{singh2020fsnet,abid2019concrete}, the probability of such a collision is negligible. However, EEG channel selection typically involves selecting from a relatively small pool of input channels. This means that the probability of different selection neurons selecting the same input channel is no longer negligible and should be addressed. To this end, we introduce a novel regularization function that couples the different selection neurons and encourages them to select distinct channels.
\newline

To validate the applicability of this method for EEG channel selection, we study its performance on two different EEG tasks: motor \insertred{execution} and auditory attention decoding (AAD). We show that this end-to-end learnable EEG channel selector achieves competitive results (better or at least equally good) compared to state-of-the-art channel selection methods for these tasks. Furthermore, the latter are often tailored or constrained to specific tasks or input feature types, while the Gumbel-softmax channel selector is widely applicable, independent of the input structure or task. It can be used for both regression and classification tasks, and can be placed behind any type of input layer, e.g. raw EEG time series or pre-computed per-channel EEG features. Our Pytorch \cite{paszke2019pytorch} implementation is available at Github\footnote{\insertred{https://github.com/AlexanderBertrandLab/Gumbel-Channel-Selection}}.
\newline

The main contributions of this paper are:
\begin{itemize}
    \item We investigate the use of the Gumbel-softmax trick in an EEG channel selector layer to learn the optimal channel selection along with the neural network weights in an end-to-end manner.
    \item We introduce a novel regularization function to avoid the occurrence of duplicate EEG channels in the selection.
    \item We demonstrate that, while being a generic (plug-and-play) approach, applicable to any EEG-related task, the Gumbel-softmax selector layer performs at least as well as state-of-the art channel selection approaches tailored towards motor \insertred{execution} and auditory match-mismatch respectively.
\end{itemize}
\ 

\section{Related Work}
\label{section: Section2}

\subsection{Channel selection}
The problem we investigate here is selecting the optimal subset $K$ channels to solve a given EEG task. This is inherently a grouped feature selection problem, with each channel containing multiple features to be selected together. In general, this can be solved with a filter, wrapper or embedded approach.
\newline

In filter approaches, the relevance of each feature in predicting the correct class label is determined using some model-independent criterion, such as, e.g., mutual information (MI). In the channel selection case however, this is complicated by the fact that each channel contains multiple features and thus requires multi-dimensional entropy estimators to determine the relevance of a single channel. One solution that has been successfully applied to EEG channel selection is applying Independent Component Analysis (ICA) to the features of each channel to transform them to new, independent features. After this transformation, their joint entropy can be estimated as the sum of their marginal entropies \cite{lan2006salient}. While this approach solves the problem of multi-dimensional entropy estimation, it still requires the computation of handcrafted features for the specific task at hand and is therefore not directly applicable in cases where raw EEG data is used directly as an input for a DNN\footnote{Note that one of the main advantages of DNNs is the fact that the network can jointly learn a classifier and a proper feature embedding in an end-to-end fashion, where only raw data is provided as an input of the network.}.
\newline

A significant drawback of filter approaches is that the channel selection is performed on a surrogate metric that is not matched to the target model that will act as the eventual classifier. To alleviate this drawback, wrapper approaches aim to find the optimal feature subset by performing a heuristic search through the space of possible subsets and evaluating the target model's performance on multiple candidate sets. For instance, Qiu et al. propose a sequential forward floating selection approach (SFFS) in combination with Common Spatial Pattern (CSP) feature extraction and Support Vector Machine (SVM) classification to select the optimal channel subset for a motor \insertred{execution} BCI problem \cite{qiu2016improved}. In each iteration of the algorithm, the channel that would improve the model's crossvalidation accuracy the most is added to the selected subset, requiring the CSP bank and SVMs to be trained multiple times. While wrapper approaches generally lead to better selections, they are expensive from a computational point of view, due to the numerous retraining of the model on a large number of feature subsets. Applying such methods when the model to be trained is a DNN would be far too time-consuming to be used in practice and will not be further discussed here.
\newline

Finally, embedded approaches jointly train the model and select the optimal features, typically by adding a regularization term to the training objective. A well-known example is the LASSO \cite{tibshirani1996regression}, which induces sparsity in a model's weights by penalizing their $L_1$-norm. LASSO can be used to perform feature selection in DNNs by driving all the weights associated with uninformative features to zero \cite{scardapane2017group} and can also be extended to select groups of features together \cite{zhao2015heterogeneous}. \insertred{However, it has been observed that LASSO has a difficult time obtaining exactly sparse solutions using standard stochastic gradient descent optimizers in the non-convex settings of neural networks. Solving this requires either using customized optimizers such as proximal gradients or manual inspection and thresholding of the weights after training \cite{scardapane2017group,bengio2012practical}}. \insertred{Another way is modeling the process of feature selection more explicitly} by using continuous relaxations of $L_0$-regularization, teaching each input neuron to either be 'on' or 'off' \cite{louizos2017learning}. \insertred{A major} downside of \insertred{all} these sparsity-inducing methods is that the amount of features selected depends on the weight of the regularization in the objective function. This means we cannot supply the model with the number of features to be selected \textit{a priori}, which makes them unfit for the given subset selection problem. If a target number of channels is to be selected, such methods have to be retrained multiple times on a trial-and-error basis. Similar to wrapper methods, this makes such methods often too time-consuming for training computationally heavy models such as DNNs.

\subsection{Discrete optimization in neural networks}

Directly solving the channel selection problem is a discrete optimization problem, while learning in neural networks is based on backpropagation, a procedure that requires a differentiable loss function and by extension, continuous parameters to be learned. However, it is possible to integrate discrete parameters in this framework by using categorical reparametrization with the Gumbel-softmax trick \cite{jang2016categorical}. This process encodes the discrete parameters as a discrete distribution to be learned. This discrete distribution is approximated by some continuous relaxation, for instance the concrete distribution \cite{maddison2016concrete}. The parameters of this distribution are then learned through standard backpropagation by employing the reparametrization trick \cite{kingma2013auto}. A general overview of this class of methods can be found in \cite{paulus2020gradient}. This framework has been used by Abid et al. to build a concrete selector layer that learns to  select $K$ features from its input \cite{abid2019concrete}. By stacking this layer on top of an autoencoder, it is possible to learn the subset of features that allow for optimal reconstruction of the complete feature set. In contrast to the embedded methods described above, this model explicitly models the amount of features to be selected and will serve as the basis for our channel selection method, as described in the next section.
\newline

\section{Proposed Channel Selection Method}
\label{section: Section 3}

\subsection{Channel selection layer}

Let $\mathcal{D} = \{(X^{(1)},y^{(1)}), (X^{(2)},y^{(2)}), \ldots , (X^{(M)},y^{(M)})\}$ be a dataset of $M$ EEG samples $X^{(i)}$ with class labels $y^{(i)}$. Each $X \in {\rm I\!R}^{N \times F}$ contains $N$ channels and $F$ features per channel. These features could be anything ranging from the raw time samples to, e.g., power features in certain frequency bands. Let $S$ indicate a subset of $K$ channels and $X_S \in {\rm I\!R}^{K \times F} $ the reduced EEG samples containing only the rows of X corresponding to the channels in $S$. Also assume we have a neural network model $f_{\theta}(X_S)$, where $\theta$ contains all the learnable parameters of the model. Our goal is then to learn the optimal $S^*$ and $\theta^*$ such that

\begin{equation}
\label{eq: loss}
    S^*, \theta^* = \arg \min\limits_{S,\theta} \mathcal{L}(f_{\theta}(X_S),y)
\end{equation}
with $\mathcal{L}(p,y)$ any loss function between the predicted label $p$ and the ground truth $y$.
\newline

 \begin{figure}[!ht]
    \centering
    \includegraphics[width = 0.5\textwidth]{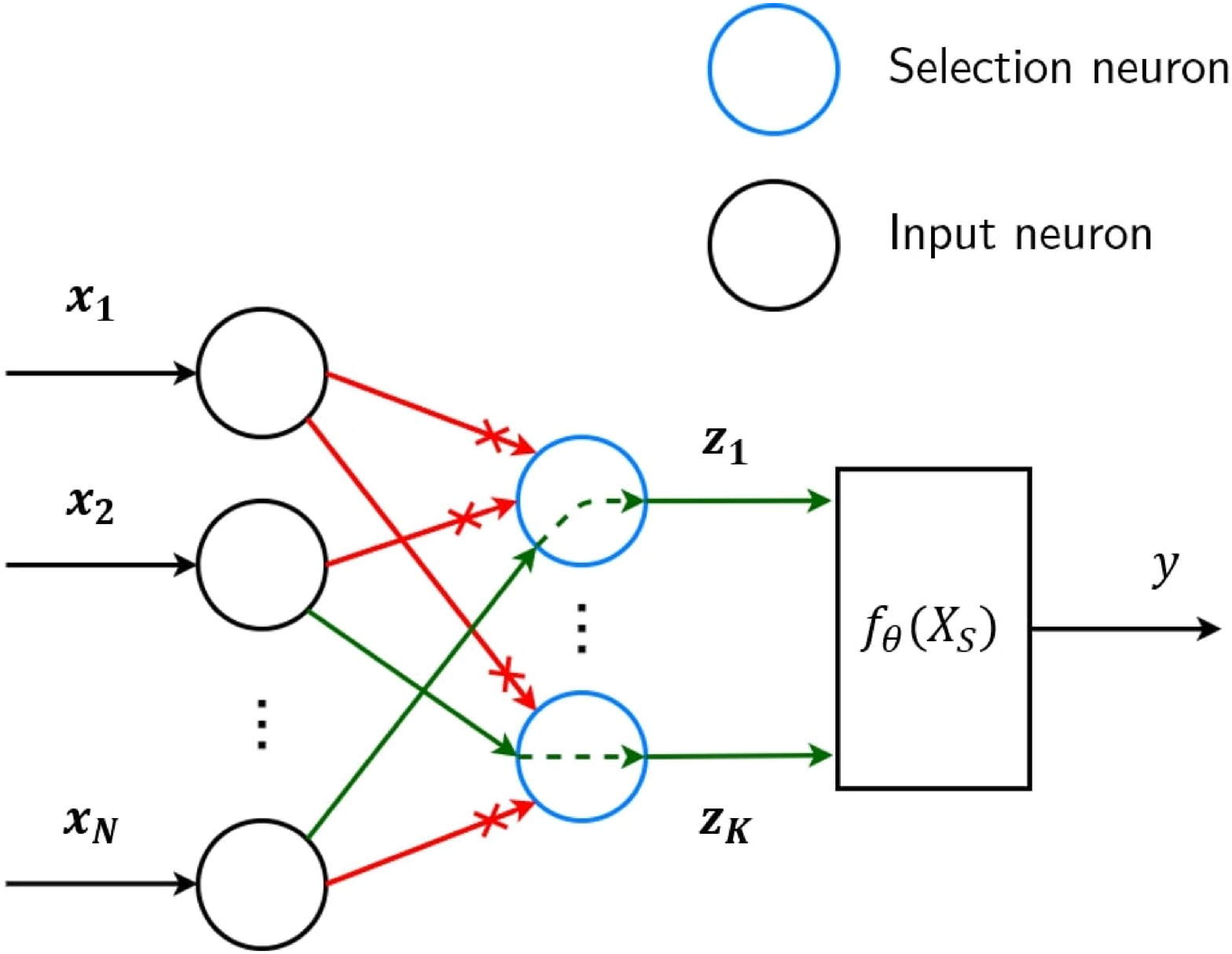}
    \caption{Illustration of the channel selection layer: $\boldsymbol{x_n}$ indicates the feature \insertred{vector} derived from channel $n$. During training, the output of each selection neuron $k$ is given by $\boldsymbol{z}_k=\boldsymbol{w}_k^{\intercal}X$, with $\boldsymbol{w}_k$ sampled from the concrete distribution \cite{maddison2016concrete}: $\boldsymbol{w}_k \sim Concrete(\boldsymbol{\alpha_k},\beta)$. \insertred{The parameters $\boldsymbol{\alpha}$ of this distribution are jointly learned with the network weights $\theta$.} At test time, only the input with the highest probability per neuron is selected (illustrated by the green arrows).}
    \label{fig: selectionlayerscheme}
\end{figure}

To accomplish this, we introduced a channel selector layer into the neural network model, \insertred{which is trained together with the rest of the network}. We propose the use of a so-called \emph{concrete} selector layer \cite{abid2019concrete}, in which $K$ selection neurons are stacked on top of each other, one for each channel to be selected (see Fig. \ref{fig: selectionlayerscheme}). Each of these selection neurons takes all channels as input and produces a single output channel. Each selection neuron is parametrized by a learnable vector $\boldsymbol{\alpha}_k \in {\rm I\!R}^{N}_{>0}$. When being fed a sample $X$, each selection neuron $k$ samples a weight vector $\boldsymbol{w}_k \in {\rm I\!R}^{N}$ from the concrete distribution \cite{maddison2016concrete}:
\begin{equation}
\label{eq: sampling}
    w_{nk} = \frac{\exp((\log \ \alpha_{nk} + G_{nk})/\beta)}{ \sum_{j=1}^{N} \exp((\log \ \alpha_{jk} + G_{jk})/\beta)}    
\end{equation}
with $G_{nk}$ independent and identically distributed (i.i.d) samples from the Gumbel distribution \cite{gumbel1948statistical} and $\beta \in (0,+\infty)$ the temperature parameter of the concrete distribution. Thus, similarly to Dropout \cite{srivastava2014dropout}, a different weight vector is sampled for each observation during training. Each neuron then computes its output channel as $\boldsymbol{z}_k=\boldsymbol{w}_k^{\intercal}X$.
\newline

Equation (\ref{eq: sampling}) can be viewed as a softmax operation, which produces weight vectors whose elements sum to one as continuous relaxations of one-hot vectors. The temperature parameter $\beta$ controls the extent of this relaxation. It can be shown that, as $\beta$ approaches 0, the distribution will become more discrete, the sampled weights will converge to one-hot vectors and the neuron will go from linearly combining to \textit{selecting} a certain input channel \cite{maddison2016concrete}. The probability $p_{nk}$ of neuron $k$ selecting a certain channel $n$ is then given by
\begin{equation}
    p_{nk}=\frac{\alpha_{nk}}{\sum_{j=1}^N\ \alpha_{jk}}\text{.}
\end{equation}
During training, the temperature parameter is decreased by an exponentially decreasing curve as in \cite{abid2019concrete}, i.e. $\beta(t) = \beta_s(\beta_e/\beta_s)^{t/T}$ with $\beta(t)$ the temperature parameter at epoch $t$, $\beta_s$ and $\beta_e$ the start and end temperature parameters and $T$ the number of epochs. This allows the network to explore various combinations of channels in the beginning of the training while forcing it to a selection operation by the end of training. At the same time, as the probability of sampling from a certain channel increases, its gradient will start to dominate the gradient of the batches, causing a positive feedback effect that drives the probability vectors $\boldsymbol{p_k}$ to (approximately) one-hot vectors as well. This means that the uncertainty of each selection neuron progressively decreases until it almost only selects one specific channel.
At test time, the stochastic nature of the network is dropped entirely and the continuous softmax is replaced by a discrete argmax. This means the weights of the neurons are replaced with fixed one-hot vectors.
\begin{equation}
\label{eq: argmax}
    w_{nk}=
    \begin{cases}
      1, & \text{if}\ n = \arg \max \limits_{j} \alpha_{jk} \\
      0, & \text{otherwise}
    \end{cases}
\end{equation}

\subsection{Duplicate channel selection}

The downside of this construction is that, since each neuron samples its weights independently, it is possible for multiple selection neurons to select the same channel, introducing redundancy in the network's input. Originally, the Gumbel-softmax selection layer was proposed for dimensionality reduction in dense auto-encoder architectures \cite{abid2019concrete} or high-dimensional input features \cite{singh2020fsnet}, where the probability of having duplicate selections is negligible. However, in the case of EEG channel selection, the probability of having a 'collision' between two selection neurons is high, as we will show in Section \ref{section: Section4}. We will refer to this problem from this point on as the duplicate channel selection problem. A straightforward (yet naive) fix for this problem would be to replace all duplicate channels with different ones, and to retrain the network weights for the newly added channels. The replacement could be done arbitrarily or one can let the network select the missing channels in a second training step from a reduced candidate pool (and repeating this process until no duplicate channels are selected anymore). However, such ad hoc fixes require additional training steps (at least one), which can be time-consuming in a context of DNNs. Furthermore, the occurrence of duplicate channels usually implies that the training process is unable to escape from a local minimum, possibly leading to a suboptimal initial channel selection, which is then carried over to all subsequent iterations. To avoid such local optima and to avoid expensive extra training phases, duplicate channels should ideally be avoided \textit{during} training instead of using post hoc (and ad hoc) fixes. To this end, we propose a regularization function that encourages the selection neurons to learn distinct channel selections within a single training phase.
\newline

To avoid this issue, we instead introduce a regularization function that encourages the selection neurons to learn distinct selections during training. Consider the selection matrix $P$, constructed by normalizing the parameter vector $\boldsymbol{\alpha}_k$ of each selection neuron and putting them in the columns of P, such that $p_{nk} = \frac{\alpha_{nk}}{\sum_{j=1}^N \alpha_{jk}}$. Thus, column $k$ of $P$ represents the probability distribution over the input channels that neuron $k$ will select as the temperature parameter goes to 0. By the end of the training procedure, the columns of this matrix approximate one-hot vectors, with the position of the 1 indicating which channel is selected for this selection neuron. It can be observed that choosing $K$ unique channels corresponds to the \textit{rows} of $P$ not containing more than a single 1-entry. During training however, the entries of the selection matrix are still continuous probabilities, so we encourage the selection neurons to pick distinct channels by penalizing the sum of the selection matrix's rows:

\begin{equation}
    \mathcal{L}(P) = \lambda \sum_{n=1}^N Relu(\sum_{k=1}^K p_{nk} - \tau)
\end{equation}
 with $\lambda$ the weight of the regularization loss, $Relu$ the rectified linear unit operation $f(x) = max(0,x)$ and $\tau$ a threshold parameter. This regularization function only applies a penalization when the sum of a channel's probabilities across the selection neurons exceeds a threshold $\tau$. Like the temperature parameter $\beta$ of the concrete distribution, this threshold is decayed exponentially during training, becoming more stringent as the distribution becomes more discrete and the selection matrix becomes more one-hot. By the end of the training, the threshold approaches 1 and duplicate channel selection is explicitly penalized. The weight $\lambda$ on the other hand controls the strength of the regularization: a higher $\lambda$ prevents more duplicate channels, but raising it too high might result in the network ignoring the original loss (\ref{eq: loss}), thereby pushing the selection process towards distinct channels, regardless of how informative they are. Important to note is that the regularization function is constructed in such a way that it does not interfere in the training at all when no duplicate channels start to arise, avoiding the introduction of unnecessary bias in the network.

\section{Materials and methods}
\label{section: Section4}

To demonstrate the generic nature of this method, we validate our method on two different EEG-BCI paradigms, \insertred{for which established neural network architectures exist: motor execution and auditory attention decoding (AAD). These two paradigms were chosen for their very different nature, both in terms of the machine learning tasks and the brain regions involved. Motor execution is a direct classification task, which typically involves the computation of spectral power in specific frequency bands of channels located in the motor cortex. On the other hand, AAD involves the detection of a stimulus-following response, which involves an implicit regression task that is generally processed in the time domain and whose most informative channels are located in the temporal lobe.  For both tasks, we will briefly review the neural network architectures employed. Note that there exist several alternative network architectures for these tasks and that the networks employed here merely serve to showcase our method, without the ambition to claim classification accuracies that go beyond the state of the art. For each task, we will also present a baseline channel selection algorithm, while refraining from using wrapper approaches due to their huge computational loads when used in conjunction with neural network models.}
 
\subsection{Motor \insertred{Execution}}
The first task we discuss is motor \insertred{execution}, a popular paradigm in the field of BCI. The goal is to decode EEG signals in motorsensory brain areas associated with imagined body movement. For our experiments, we make use of the \insertred{publicly available} High Gamma Dataset\footnote{https://github.com/robintibor/high-gamma-dataset} \cite{schirrmeister2017deep}. This dataset contains 128-channel EEG recordings from 14 subjects, with about 1000 trials of executed movement per subject following a visual cue. These movements belong to one of four classes: left hand, right hand, feet and rest. We employ the same preprocessing procedure as \cite{schirrmeister2017deep}, that is, resampling the data to 250 Hz, highpass-filtering the data at 4 Hz, standardizing the data for each electrode and epoching the data in 4.5 second segments, taking the 0.5 seconds before each visual cue and the 4 seconds after. \insertred{80\% of the training set is used for training and 20\% kept for validation}. We report mean test accuracy across the subjects on a held-out test set of about 180 trials per subject. 
\newline

We classify the filtered time series with the parallel multiscale filter bank convolutional neural network (MSFBCNN) proposed in \cite{wu2019parallel}. \insertred{This network first passes the input data through 4 temporal convolution layers of increasing kernel sizes in parallel. The outputs are then concatenated and fused through a layer of spatial filters. Temporal average pooling combined with square and log non-linearities then yield band power related features, which are classified by a dense layer. While the full architecture of the network is described in \cite{wu2019parallel}, we provide a detailed summary of the network in table format in Appendix A for the sake of completeness and reproducibility of our results. The size of this network varies from 4744 parameters (for 1-channel input) to 24 344 parameters (for 50-channel input, the maximum amount we will use in our experiments). When selecting $K$ from $N$ channels, the concrete selection layer adds $NK$ parameters, thus ranging from 128 to 6400 parameters.}
\newline

For motor \insertred{execution}, we compare the performance of the proposed Gumbel-softmax method with the mutual information (MI) based channel selection approach described in \cite{lan2006salient}. In short, it can be described as follows: the joint MI \insertred{between} each (per-channel) block of features and the class labels is computed and the channel with the maximal MI is added to the set of selected channels. Then, the joint MI \insertred{between the class labels and the set of previously selected channels combined with each remaining channel (separately)} is computed and the channel that maximizes this value is added to the selected set. This process is repeated until the desired number of channels are selected. Since we first need to craft informative features to perform this procedure, we compute the spectral power of 9 frequency bands between 4 and 40 Hz for each channel, \insertred{each with a bandwidth of 4 Hz}. These features were previously employed in filter bank CSPs for motor \insertred{execution} \cite{ang2008filter}. Note that the feature extraction is only used here to inform the MI channel selection procedure, whereas the classification by the MSFBCNN is performed directly on the filtered time series.
\newline

\subsection{Auditory match-mismatch}
A second, entirely different task we discuss here is speech decoding, more specifically, the auditory match-mismatch paradigm \cite{wong2018accurate}. Given two candidate speech envelopes and an EEG recording, the goal is to classify which of the envelopes actually elicited the EEG response and which one is an `imposter'. We employ the dataset described in \cite{monesi2020lstm}, containing \insertred{64-channel EEG recordings from 48 normal hearing subjects listening to narrated audiobooks, yielding about 83 hours of EEG and corresponding audio recordings in total. From these recordings, 10-second windows with 90\% overlap are extracted. The envelope of the matching speech stimulus is then estimated using the gamma-tone filterbank method from \cite{biesmans2016auditory}. A similar imposter envelope is then generated by taking the 10-second window that starts 1 second after the real one has ended. 80\% of the data is used for training, 10\% for validation and 10\% for testing.} As before, we report mean test accuracy across the subjects on this held-out test set. 
\newline

For classification, we follow the approach of \cite{accou2020dilated}, using their dilated convolutional neural network (DCNN) model. Similar to the previous network, this model uses the raw EEG traces as inputs and therefore does not require a prior feature construction step. \insertred{In this architecture, the EEG is first passed through a spatial convolution layer and then through a series of dilated time convolution layers with an increasing dilation factor, maximizing the size of the receptive field. The goal of these layers is to extract the stimulus-following response from the EEG, which is known to be correlated to the energy envelope of the attended speech. The two speech envelopes are passed through similar dilated time convolutions, after which each envelope is compared with the EEG by cosine similarity. These similarity scores are finally passed through a dense layer for classification. The full architecture and motivations therefore are described in \cite{accou2020dilated}. We provide a detailed summary of the network in table format in Appendix B for the sake of completeness and reproducibility of our results. The size of this network varies from 4 193 parameters (for 1-channel input) to 8473 parameters (for 64-channel input). When selecting $K$ from $N$ channels, the concrete selection layer adds $NK$ parameters, thus ranging from 64 to 4096 parameters.}
\newline

Unfortunately, using MI as a channel selection benchmark for this application is not straightforward, as MI requires a set of discriminative features to work on. For motor \insertred{execution}, it is generally known that the power in certain spectral bands can indeed serve as a discriminative feature. In the field of AAD on the other hand, there is no clear set of EEG/speech features that perform the same role. Instead, we employ a greedy channel selection procedure based on the least-squares utility metric as described in \cite{narayanan2019analysis}, which is currently the state-of-the-art channel selection method in EEG-based speech decoding paradigms \cite{narayanan2020optimal}. In this setting, a linear decoder is trained to reconstruct the matching speech stimulus from the EEG signal, which constitutes a least-squares (LS) regression problem. The utility metric of a channel is then defined as the increase in LS-cost when this channel is dropped from the regression and the regression parameters are re-optimized, which can be calculated very efficiently as shown in \cite{bertrand2018utility}. We can use this to select $K$ channels by iteratively removing the channels with the lowest utility metric until there are $K$ channels left.

\subsection{Training procedure}
For both tasks, we build the network for Gumbel-softmax channel selection by inserting the channel selection layer between the input layer and the baseline networks as described above. Using the data of all subjects simultaneously, we jointly train the selection layer and the network weights using the Adam optimizer (with a learning rate of $0.001$) \cite{kingma2014adam}. \insertred{Note that this corresponds to a subject-independent channel selection}. During training, the temperature parameter $\beta$ is decayed from 10 to 0.1 and the regularization threshold $\tau$ from 3 to 1.1. For the regularization weight $\lambda$ we chose a value of $0.1$, which worked well for our applications, where the supervised loss was typically between $1$ and $0.1$. For applications where the supervised loss is higher or lower than this, $\lambda$ should be scaled accordingly. We track the convergence of the channel selection by analyzing the normalized entropy of the distribution of each selection neuron, computed as:
\begin{equation}
    H(\boldsymbol{\alpha_k}) = -\frac{1}{\log N} \sum_{j=1}^N \alpha_{jk} \log(\alpha_{jk}).
\end{equation}
When the mean entropy of all selection neurons drops below a predefined threshold (we took $0.05$ in our experiments) \insertred{we activate an early stopping procedure, continuing training until the validation loss stops to decrease} or the maximum amount of epochs is reached (150 for the motor \insertred{execution} and 50 for the auditory match-mismatch task), we consider the selection process to have converged and the training finished. At that point, the parameters of the channel selection layer are frozen and the layer operates in its deterministic selection mode for evaluation (see section \ref{section: Section 3}). We performed 10 runs of each training session and report results across all runs through box plots. \insertred{These are tested for statistical significance through independent samples t-tests.}

\begin{figure}[htbp]
	\begin{minipage}{0.99\linewidth}
		\includegraphics[trim={1cm 0 0 0},width = \textwidth,clip]{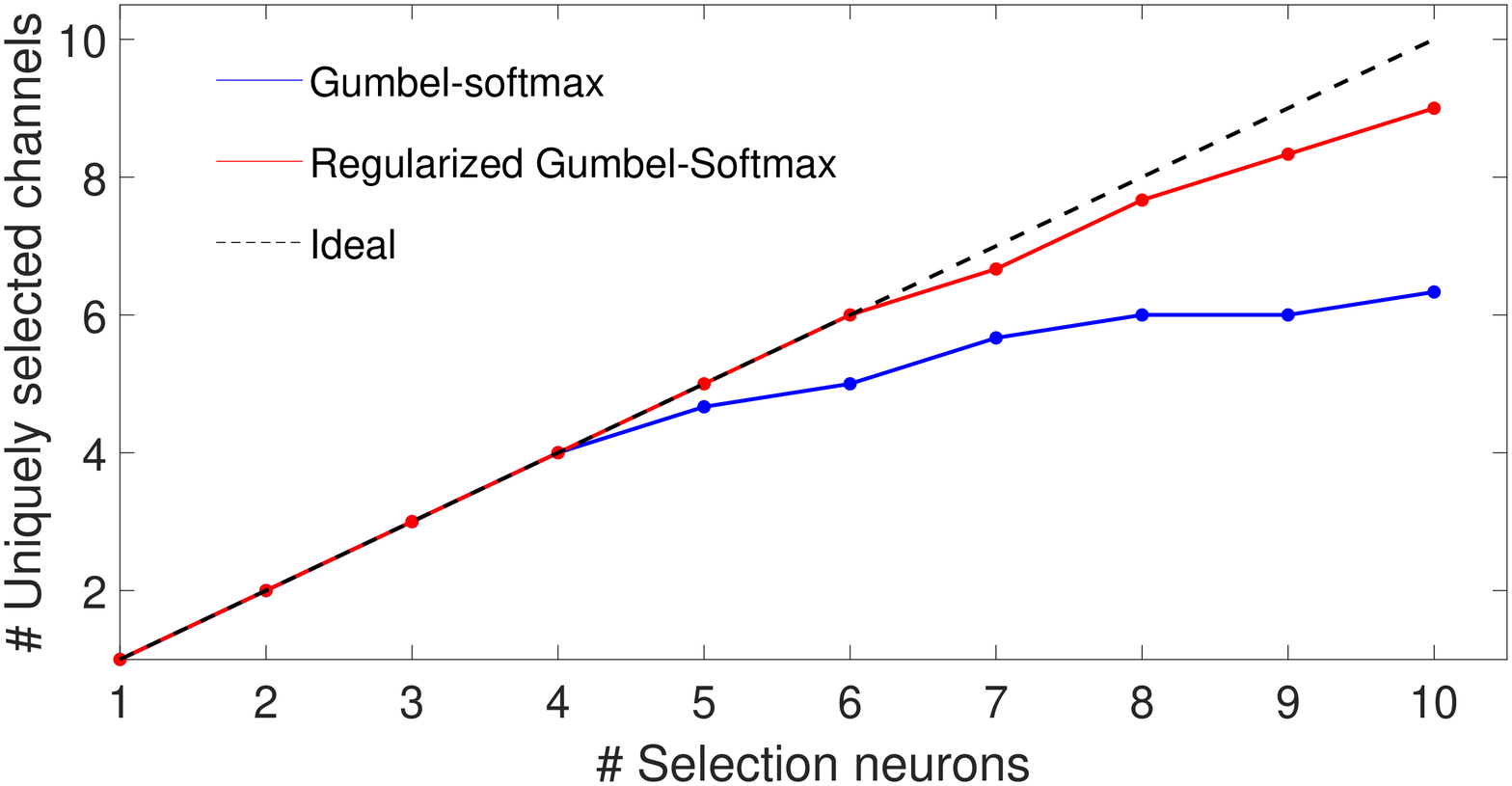}
		\caption*{(a)}
	\end{minipage}%
	\hfill
	\begin{minipage}{0.99\linewidth}
		\includegraphics[trim={1cm 0 0 0},width = \textwidth,clip]{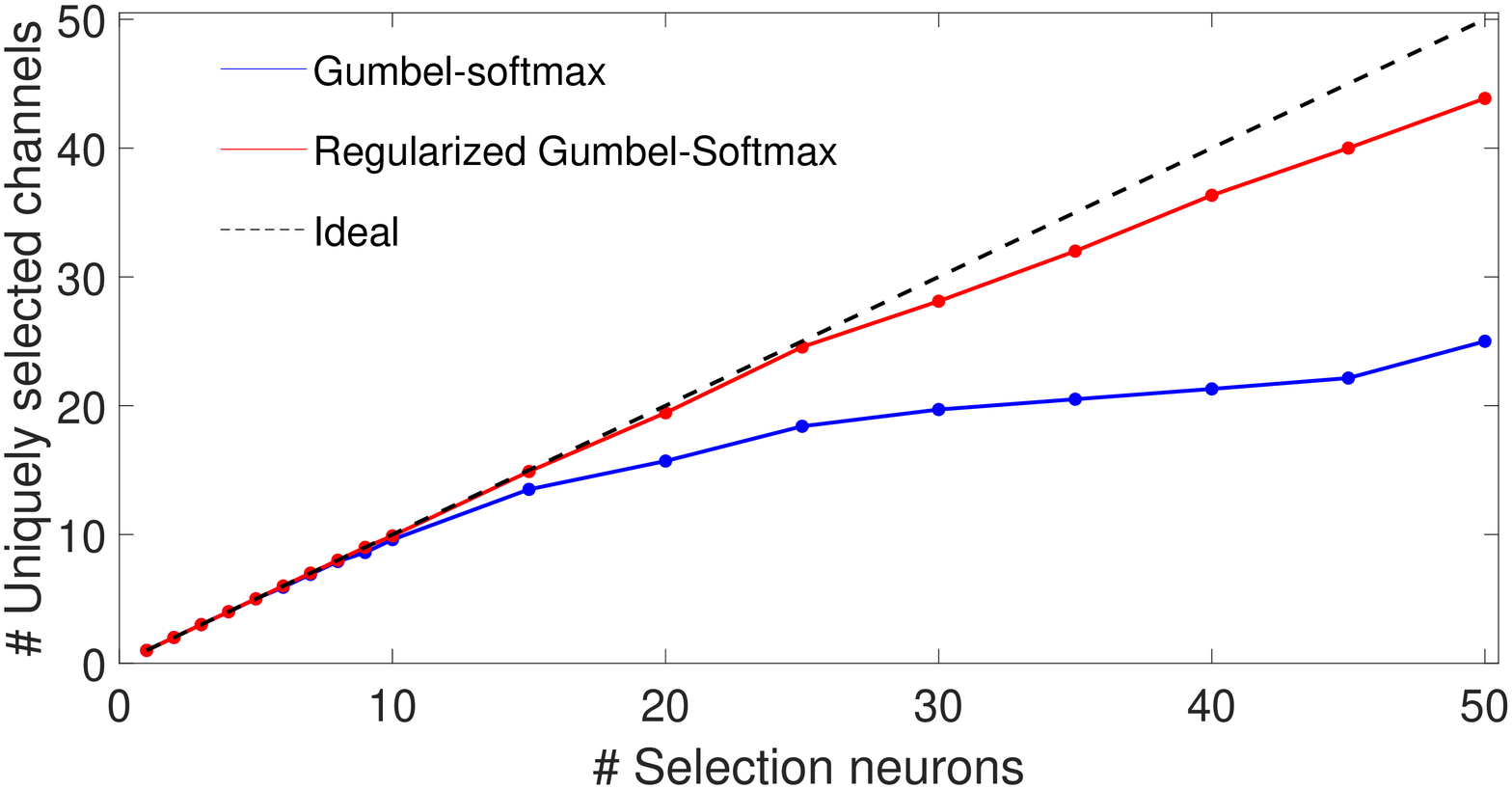}
		\caption*{(b)}
	\end{minipage}%
	\hfill
	\begin{minipage}{0.99\linewidth}
        \includegraphics[trim={1cm 0 0 0},width = \textwidth,clip]{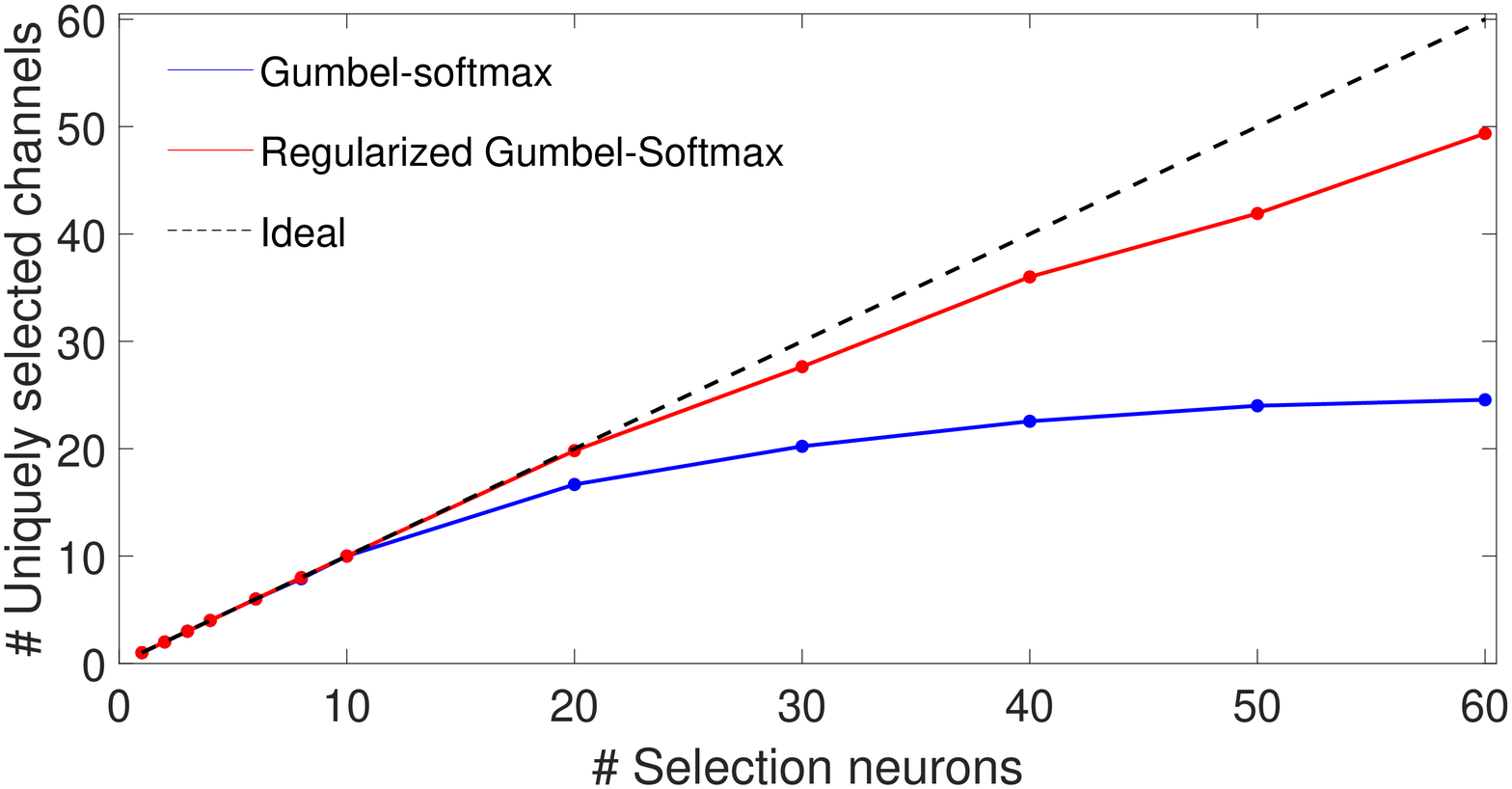}
		\caption*{(c)}
	\end{minipage}%
	\hfill
	\caption{\insertred{Comparison of the amount of duplicate channels occurring in the Gumbel-softmax and regularized Gumbel-softmax method in function of the amount of channels to be selected (a) for a 10-channel subset of the motor execution task, (b) for the full motor execution task and (c) for the auditory match-mismatch task. The displayed curves are averaged over 10 runs}}
	\label{fig: performance_gumbeluniqueness}
\end{figure}

\section{Experimental Results}
\label{section: Section5}

\subsection{Duplicate channel selection}

\begin{figure*}[!htbp]
    \centering
    \includegraphics[trim={1.6cm 0 -1cm 0}, height=0.4\textheight, clip]{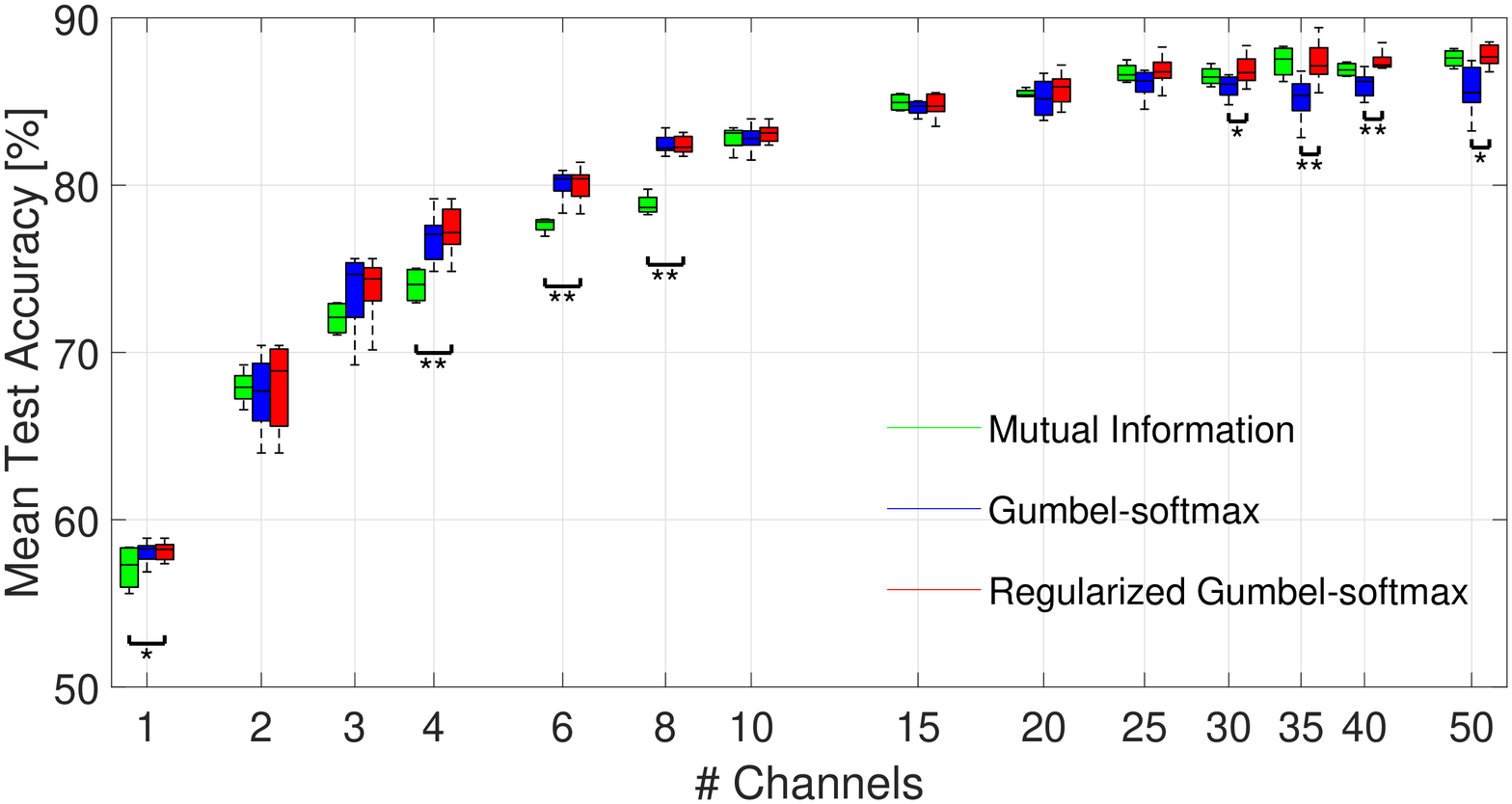}
    \caption{Comparison of the Gumbel-softmax channel selection strategies on the motor \insertred{execution} dataset. Mean test accuracies across the subjects are plotted as a function of the number of selection neurons. Note that duplicate channels can occur in the Gumbel-softmax methods, so the x-axis represents the \textit{maximum} number of channels the algorithm can select.  The displayed boxplots are computed over 10 runs and compared with independent samples t-test (no correction for multiple comparison). $*$ indicates statistically significant difference with $p<0.05$, $**$ with $p <0.005$.}
    \label{fig: motorimageryperformance}
\end{figure*}

\begin{figure*}[!htbp]
    \centering
    \includegraphics[trim={1.6cm 0 -1cm 0}, height =0.4\textheight, clip]{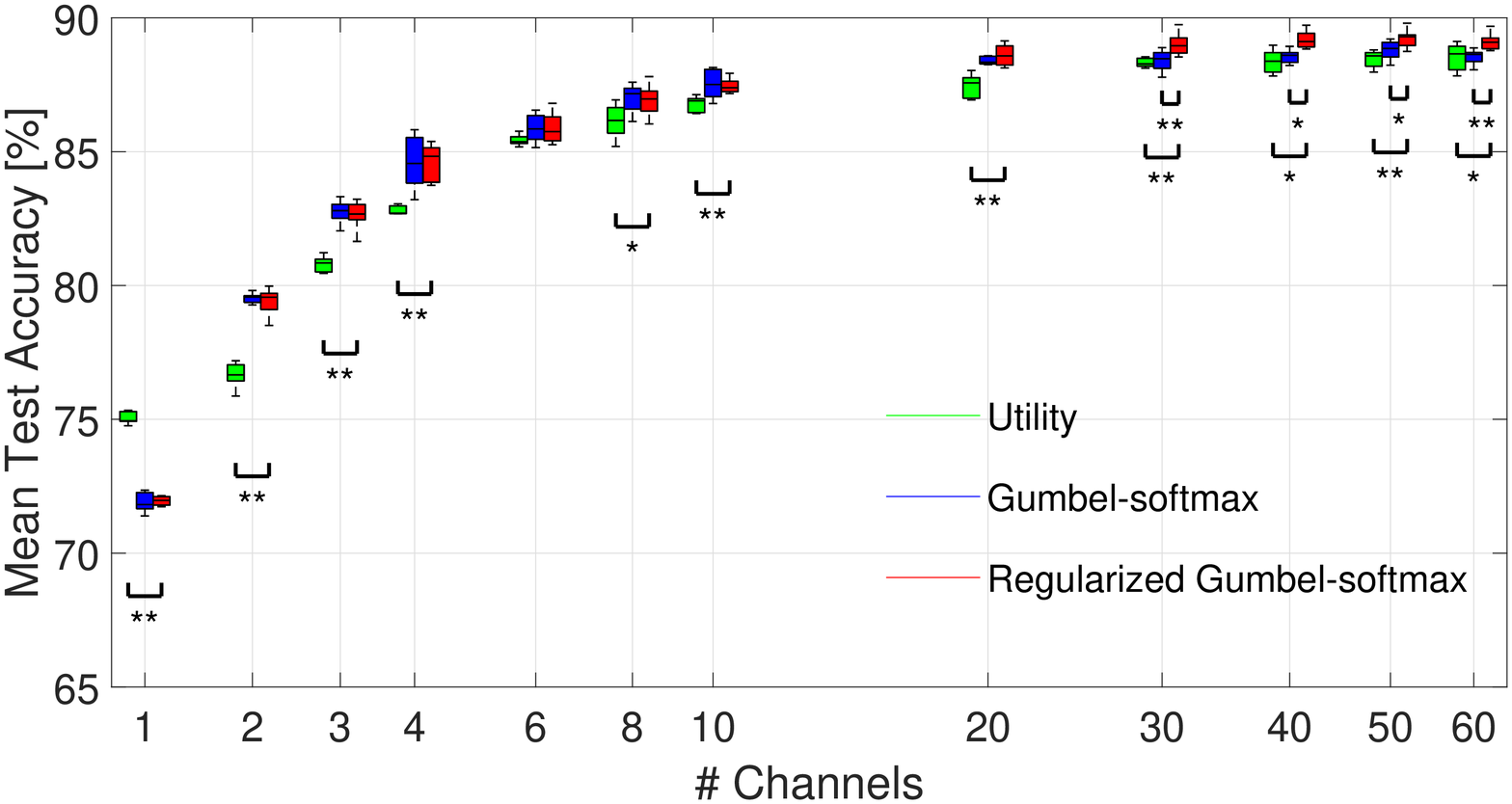}
    \caption{Comparison of the channel selection strategies on the auditory match-mismatch dataset. Mean test accuracies across the subjects are plotted as a function the number of selection neurons.  The displayed boxplots are computed over 10 runs  and compared with independent samples t-test (no correction for multiple comparison). $*$ indicates statistically significant difference with $p<0.05$, $**$ with $p <0.005$.}
    \label{fig: AADperformance}
\end{figure*}

We first study the effect of our proposed regularization to avoid duplicate channel selection. \insertred{To demonstrate the higher probability of duplicate channels when selecting from a smaller pool of channels, we first apply the selector layer to a small subset from the motor execution dataset (the 10 highest-ranking channels yielded by the MI algorithm). Fig. \ref{fig: performance_gumbeluniqueness}a shows the number of uniquely selected channels in function of the amount of selection neurons for this experiment. Without regularization, duplicates already start occurring when only selecting 5 out of 10 channels, which is clearly undesired. Adding the regularization term steers the network away from these suboptimal solutions, postponing the point at which these duplicates occur. Fig. \ref{fig: performance_gumbeluniqueness}b shows the same analysis for the full motor execution dataset, while the corresponding test accuracy is shown in Fig. \ref{fig: motorimageryperformance}}. Here, duplicates can start appearing when selecting 10 out of \insertred{128} channels, while being quite common from $K=15$ on. \insertred{Starting from $K=25$}, a performance gap between normal and regularized Gumbel-softmax can be seen in Fig. \ref{fig: motorimageryperformance}, implying that, without the regularization, the algorithm gets stuck in a local minimum and converges to a suboptimal performance. \insertred{ Figures \ref{fig: performance_gumbeluniqueness}c and \ref{fig: AADperformance} show the same analysis for the auditory match-mismatch task, with the duplicates starting to appear between $K=10$ and $K=20$ and the performance gap at $K=30$.}

\begin{figure*}[!htbp]
	\begin{minipage}{0.45\linewidth}
		\includegraphics[trim={10cm 0cm 10cm 0cm},clip,width=\textwidth]{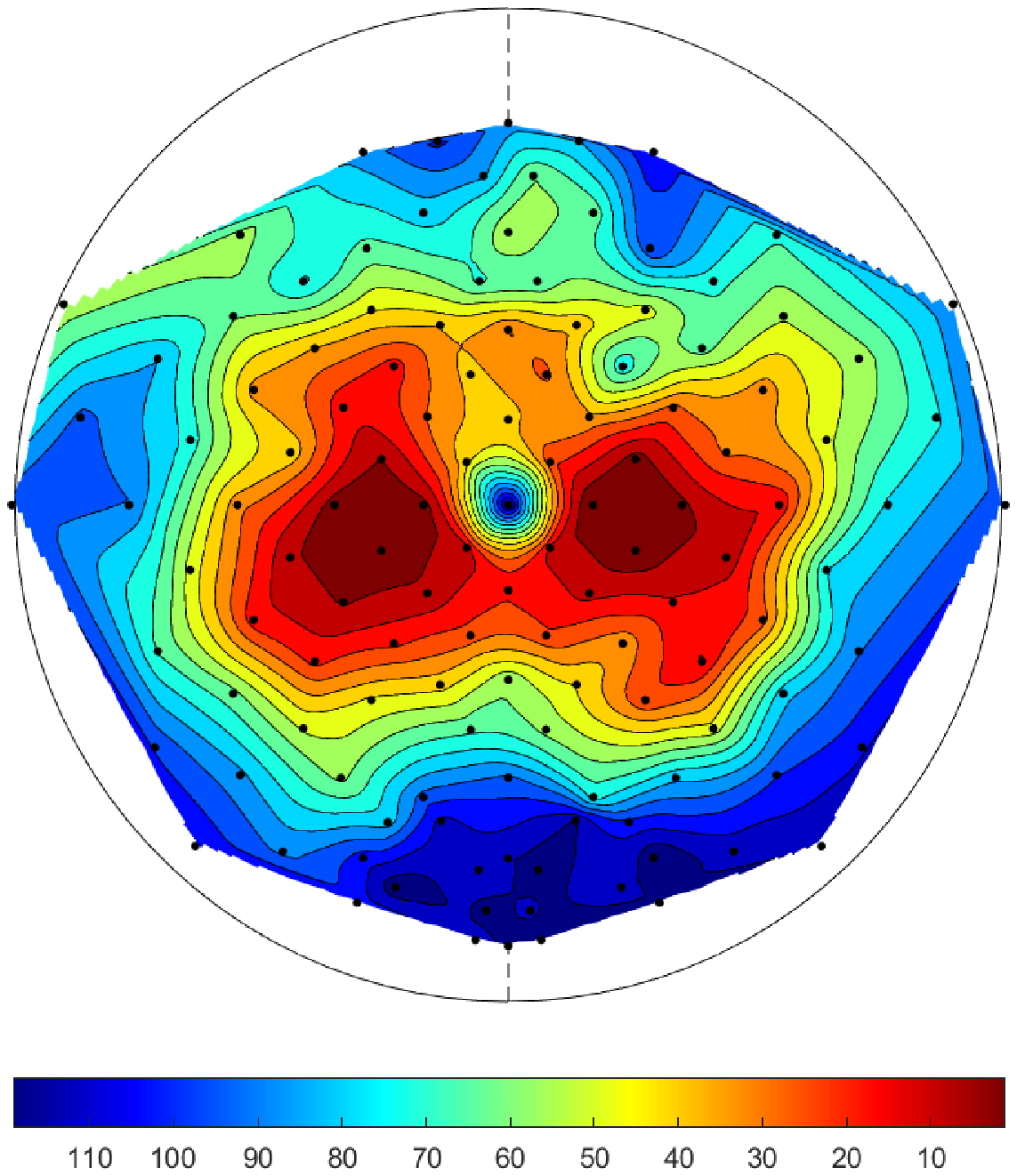}
		\caption*{(a)}
		\label{fig:figure1}
	\end{minipage}%
	\hfill
	\begin{minipage}{0.45\linewidth}
		\includegraphics[trim={10cm 0cm 10cm 0cm},clip,width=\textwidth]{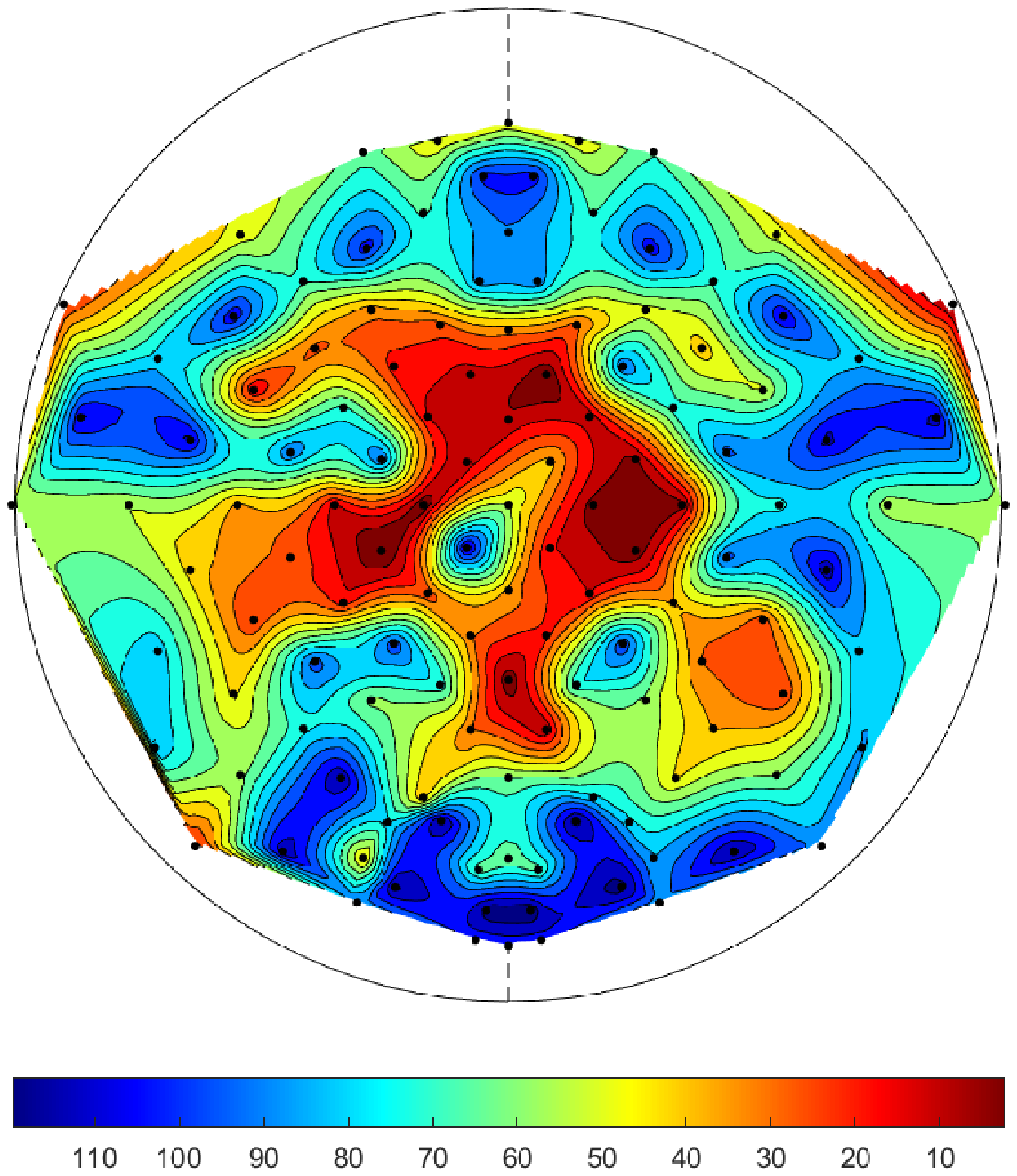}
		\caption*{(b)}
		\label{fig:figure3}
	\end{minipage}
	\hfill
	\caption{\insertred{Topoplots showing the channel ranking resulting from the different channel selection strategies for the motor execution dataset. (a):  Mutual information (b): Regularized Gumbel-softmaxs. A ranking of 1 corresponds to the best channel. For the Gumbel-softmax method, the ranking is obtained by sorting the channels by their probability of being selected across 10 runs. }}
	\label{fig: topoplots_HGD}
\end{figure*}

\begin{figure*}[!htbp]
	\begin{minipage}{0.45\linewidth}
		\includegraphics[trim={10cm 0cm 10cm 0cm},clip,width=\textwidth]{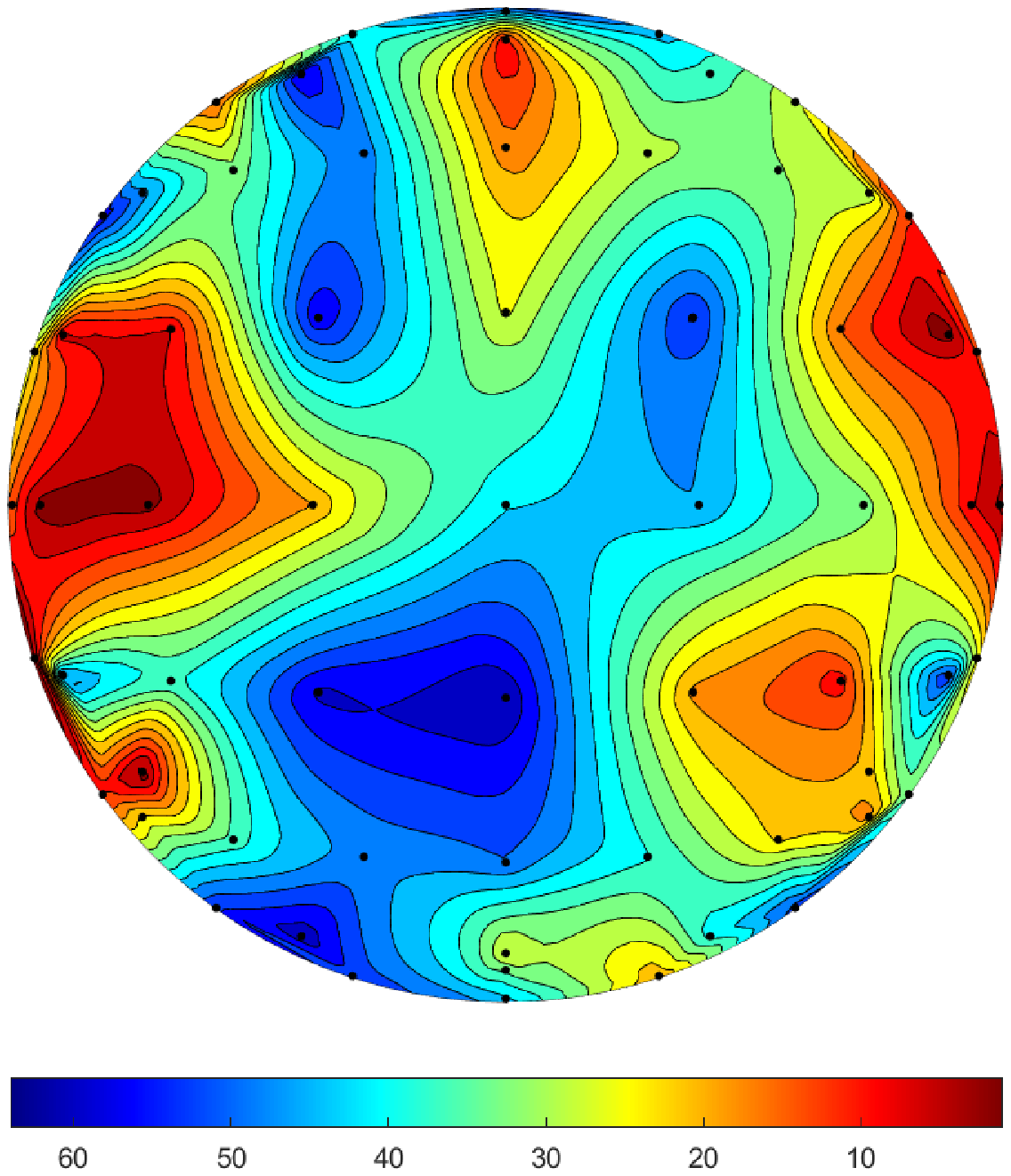}
		\caption*{(a)}
		\label{fig:figure1}
	\end{minipage}%
	\hfill
	\begin{minipage}{0.45\linewidth}
		\includegraphics[trim={10cm 0cm 10cm 0cm},clip,width=\textwidth]{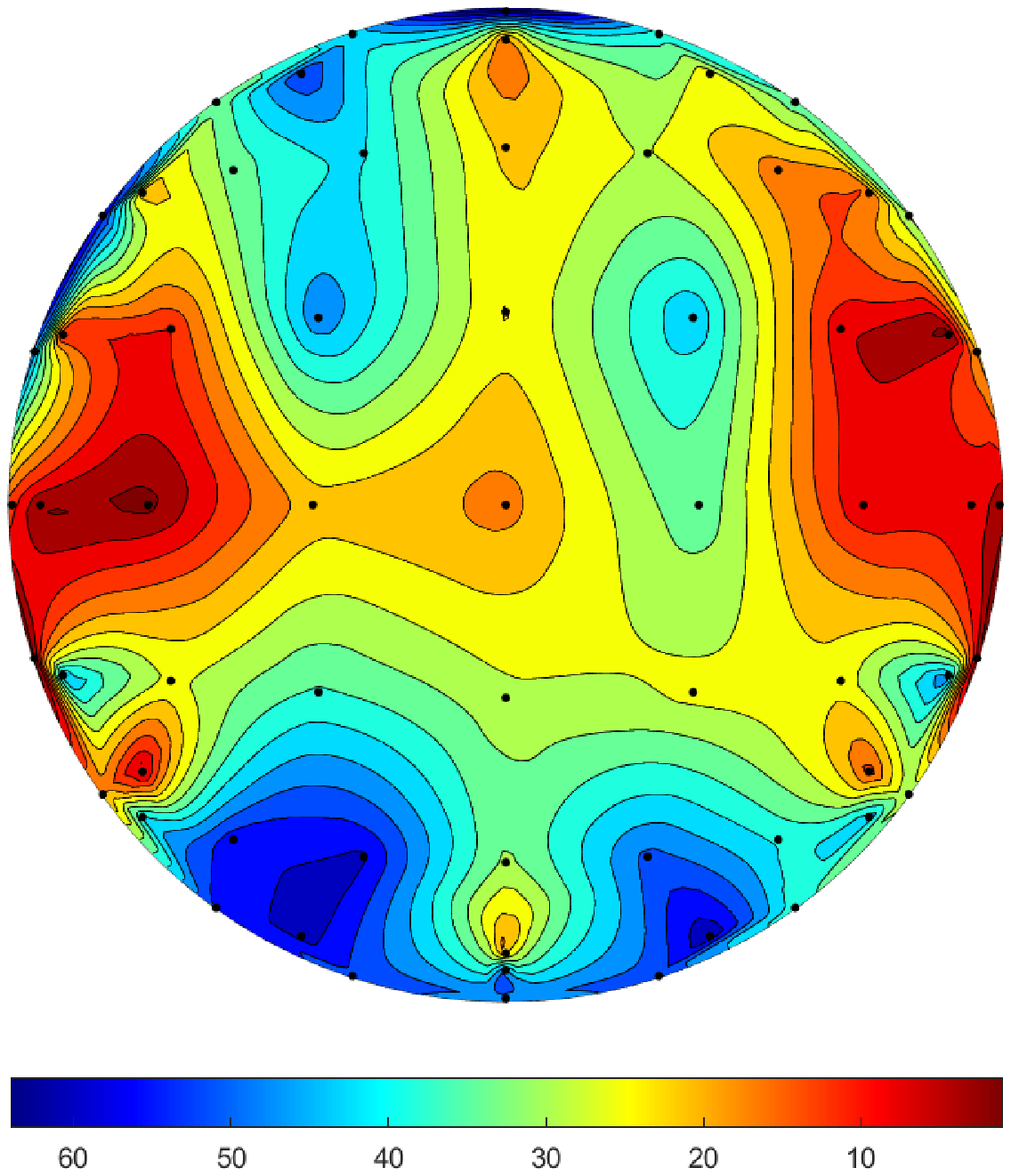}
		\caption*{(b)}
		\label{fig:figure2}
	\end{minipage}%
	\hfill
	\caption{\insertred{Topoplots showing the channel ranking resulting from the different channel selection strategies for the auditory match-mismatch task. (a): Utility metric. (b): Regularized Gumbel-softmax. A ranking of 1 corresponds to the best channel. For the Gumbel-softmax method, the ranking is obtained by sorting the channels by their probability of being selected across 10 runs.}}
	\label{fig: topoplots_AAD}
\end{figure*}

\subsection{Channel selection task}

\insertred{In our next experiments, we study the channels selected and the accuracies reached by the Gumbel-softmax and baseline channel selection strategies. Note that our goal here is not exceeding existing channel selection strategies by a statistically significant amount, but rather showing that Gumbel-softmax, being a versatile and generic `plug and play' method, can achieve at least comparable performance as less generic, task-dependent selection methods. Results on statistical significance are added merely for the sake of completeness (and without multiple comparisons correction due to the large number of tests).}
\newline

\subsubsection{Motor \insertred{execution} task}

Fig. \ref{fig: motorimageryperformance} compares the test accuracies reached by the regularized Gumbel-softmax and the MI algorithm. \insertred{We limited the amount channels to be selected to 50 out of 128, since no significant performance gains were observed beyond this point.} The Gumbel-softmax method performs \insertred{at least on par with (and often better than)} the MI algorithm, while having two main advantages over it. Firstly, the MI channel selection method is computationally expensive, performing multiple iterations of ICA's of growing size to perform the channel selection and then still needing the network to be trained on the selected channel subset. Secondly, the MI algorithm requires the computation of handcrafted features since computing accurate entropies of the raw time series would pose a far too high-dimensional problem. The Gumbel-softmax method on the other hand is able to perform the channel selection on the raw time series and jointly learns the network weights alongside the optimal selection. Also interesting to note is that at only \insertred{8} channels, the network's performance differs only about \insertred{5\%} from the \insertred{maximal} performance, showing that deep neural network models can still be powerful tools in the low-channel settings of mobile EEG. \insertred{Topoplots of the channels selected by both algorithms are shown in Fig. \ref{fig: topoplots_HGD}. From these, it can be concluded that the selection layer indeed consistently selects most of the channels near the motor cortex, as expected.}
\newline

\subsubsection{Auditory match-mismatch task}

The results of the auditory match-mismatch task are shown in Fig. \ref{fig: AADperformance}. We observe that the Gumbel-softmax in most cases performs at least as well as \insertred{(and often better than)} the utility metric despite the former being a more generic method. \insertred{One surprising downside of the Gumbel-softmax is its lower performance (compared to the baseline) in the case where $K=1$. One hypothesis is that greedy or ranking-based methods are typically optimal in this setting, as they have the luxury of testing each channel one by one and selecting the best one. Once more channels need to be selected, these methods struggle to take the interactions across channels into account and start to underperform compared to the Gumbel-softmax. Nevertheless, even in the case of $K=1$, the Gumbel-softmax is able to select a channel with a reasonably good performance.} Finally, high accuracies compared to the full channel baseline can already be achieved with a low amount of channels, with 10 channels already achieving an accuracy close to that of the full-channel baseline. \insertred{Topoplots of the selected channels are displayed in Fig. \ref{fig: topoplots_AAD}. As expected, both methods primarily select channels near the left and right temporal lobe, the location of the auditory cortex and a known region of interest for auditory attention tasks \cite{mirkovic2015decoding}. Both topoplots look very alike, suggesting that the stronger performance of the Gumbel-softmax does not stem from finding a particularly informative channel that the utility metric does not. Rather, it appears that the Gumbel-softmax is better at combining channels that yield a higher performance together, something greedy selection strategies have more trouble with.}

\section{Conclusion and future outlook}
\label{section: Section6}

We have proposed the use a concrete selector layer \cite{abid2019concrete}, to solve the EEG channel selection problem in an end-to-end fashion. This method embeds the channel selection as part of the training of the model, dealing with its discrete nature by employing categorical reparametrization with Gumbel-softmax. We have demonstrated that directly applying this method to the task of EEG channel selection results in redundant selections containing duplicate channels. We addressed this issue by introducing a novel regularization function that encourages the selection neurons to choose distinct channels, which was shown to increase the performance of the algorithm. 
\newline 

The performance of this method has been evaluated on two different EEG tasks, motor \insertred{execution} and auditory match-mismatch. On both these tasks, the experimental results indicate that the Gumbel-softmax performs at least as well as (and often better than) state-of-the-art methods that were tailored for these specific tasks: mutual information (for motor \insertred{execution}) and greedy channel selection with the utility metric (for auditory match-mismatch). Important to note here is that both these benchmark algorithms are not easily usable on the other task. MI requires the computation of task-specific, per-channel features to compute the mutual information with the class label with, features that are not readily available in the case of auditory match-mismatch. The utility metric requires the problem at hand to be formulated as a least-squares regression problem, which is not possible in the case of motor \insertred{execution}. The Gumbel-softmax method on the other hand is a very general method and can be readily applied to any EEG regression or classification task, whether the inputs are pre-computed features or raw time series.
\newline

A second advantage of the Gumbel-softmax approach is that it is an embedded method that jointly learns the optimal channel selection and the weights of the neural network classifier/regressor model. In contrast, traditional filter methods require a separate channel selection step and a subsequent training of the model on the selected subset, and therefore the channel selection strategy is not necessarily matched to the target model that will eventually be used for classification. On the other hand, wrapper methods require the model to be trained multiple times, a computationally infeasible demand when dealing with neural networks.
\newline

Finally, the Gumbel-softmax method has a great plug-and-play value: applying it to an existing model simply requires putting a channel selection layer in front of it, given that the input layer of the neural network can be scaled to accommodate the lower input dimensionality of the required number of channels. Additionally, our use of a regularization function for distinct selections can be extended with additional constraints on the channels to be selected. For example, one possibility is selecting the channels that not only optimize performance, but also minimize the inter-electrode distance as much as possible, as is required in the design of miniaturized EEG sensor networks \cite{narayanan2019analysis, narayanan2020optimal}.

\section*{Acknowledgements}

We would like to thank Professor Tom Francart, Bernd Accou and Mohammad Jalilpour Monesi for contributing the dataset for the auditory match-mismatch task.

\bibliographystyle{IEEEtran}
\bibliography{mybib.bib}

\onecolumn

\appendices

\section{MSFBCNN architecture}

\begin{table*}[!h]
\begin{tabular}{llllllll}
\hline
\textbf{Layer} & \textbf{\# Filters} & \textbf{Kernel} & \textbf{Stride} & \textbf{\# Params} & \textbf{Output} & \textbf{Activation} & \textbf{Padding} \\ \hline
Input          &                     &                 &                 &                    & (C,T)           &                     &                  \\
Reshape        &                     &                 &                 &                    & $(1,T,C)$       &                     &                  \\
Timeconv1      & $F_T$               & $(64,1)$        & $(1,1)$         & $64F_T$            & $(F_T,T,C)$     & Linear              & Same             \\
Timeconv2      & $F_T$               & $(40,1)$        & $(1,1)$         & $40F_T$            & $(F_T,T,C)$     & Linear              & Same             \\
Timeconv3      & $F_T$               & $(26,1)$        & $(1,1)$         & $26F_T$            & $(F_T,T,C)$     & Linear              & Same             \\
Timeconv4      & $F_T$               & $(16,1)$        & $(1,1)$         & $16F_T$            & $(F_T,T,C)$     & Linear              & Same             \\
Concatenate    &                     &                 &                 &                    & $(4F_T,T,C)$    &                     &                  \\
BatchNorm      &                     &                 &                 & $2F_T$             & $(4F_T,T,C)$    &                     &                  \\
Spatialconv    & $F_S$               & $(1,C)$         & $(1,1)$         & $4CF_TF_S$         & $(F_S,T,1)$     & Linear              & Valid            \\
BatchNorm      &                     &                 &                 & $2F_S$             & $(F_S,T,1)$     &                     &                  \\
Non-linear     &                     &                 &                 &                    & $(F_S,T,1)$     & Square              &                  \\
AveragePool    &                     & $(75,1)$        & $(15,1)$        &                    & $(F_S,T/15,1)$  &                     & Valid            \\
Non-linear     &                     &                 &                 &                    & $(F_S,T/15,1)$  & Log                 &                  \\
Dropout        &                     &                 &                 &                    & $(F_S,T/15,1)$  &                     &                  \\
Dense          & $N_C$               & $(T/15,1)$      & $(1,1)$         & $F_S(T/15)N_C$     & $N_C$           & Linear              & Valid            \\ \hline
\end{tabular}
\caption{\insertred{Architecture of the MSFBCNN used for motor execution classification. In the model we use $T=1125, F_T=10, F_S=10$ and $N_C=4$. The input is provided by the output of the selection layer, meaning $C$=$K$, the number of channels to be selected. This table is cited from \cite{wu2019parallel}.}}
\end{table*}

\section{DCNN architecture}
\begin{table*}[!h]
\begin{tabular}{lllllllll}
\hline
\textbf{Layer} & \textbf{\# Filters} & \textbf{Kernel} & \textbf{Stride} & \textbf{Dilation} & \textbf{\# Params} & \textbf{Output} & \textbf{Activation} & \textbf{Padding} \\ \hline
EEG Input      &                     &                 &                 &                   &                    & (1,T,C)         &                     &                  \\
Spatialconv    & $F_S$               & $(1,C)$         & $(1,1)$         & $(1,1)$           & $CF_S$             & $(F_S,T,1)$     & Linear              & Same             \\
Timeconv1      & $F_T$               & $(3,1)$         & $(1,1)$         & $(1,1)$           & $3F_TF_S$          & $(F_T,T,1)$     & Relu                & Same             \\
TimeConv2      & $F_T$               & $(3,1)$         & $(1,1)$         & $(3,1)$           & $3F_TF_T$          & $(F_T,T,1)$     & Relu                & Same             \\
TimeConv3      & $F_T$               & $(3,1)$         & $(1,1)$         & $(9,1)$           & $3F_TF_T$          & $(F_T,T,1)$     & Relu                & Same             \\
Envelope Input 1    &                     &                 &                 &                   &                    & $(1,T,1)$       &                     &                  \\
Envelope Input 2    &                     &                 &                 &                   &                    & $(1,T,1)$       &                     &                  \\
Timeconv4      & $F_T$               & $(3,1)$         & $(1,1)$         & $(1,1)$           & $3F_T$             & $(F_T,T,1)$     & Relu                & Same             \\
Timeconv5      & $F_T$               & $(3,1)$         & $(1,1)$         & $(3,1)$           & $3F_T$             & $(F_T,T,1)$     & Relu                & Same             \\
Timeconv6      & $F_T$               & $(3,1)$         & $(1,1)$         & $(9,1)$           & $3F_T$             & $(F_T,T,1)$     & Relu                & Same             \\
Dot EEG-Env1   &                     &                 &                 &                   &                    & $(F_T,F_T)$     &                     &                  \\
Dot EEG-Env2   &                     &                 &                 &                   &                    & $(F_T,F_T)$     &                     &                  \\
Concatenate    &                     &                 &                 &                   &                    & $(2F_T,F_T)$    &                     &                  \\
Flatten        &                     &                 &                 &                   &                    & $2F_TF_T$      &                     &                  \\
Dense          & $1$                 & $2F_TF_T$      & $1$             &                   &                    & $1$             &                     &                  \\ \hline
\end{tabular}
\caption{\insertred{Architecture of the DCNN used for auditory match-mismatch  classification. In the model we use $T=640, F_S=8, F_T=16$. The input is provided by the output of the selection layer, meaning $C$=$K$, the number of channels to be selected. Parameters of Timeconv4-6 are shared for both envelope inputs. Dot indicates a normalized dot product in the time dimension, for each feature map computed for the EEG and for each envelope. For more details on this architecture, see \cite{accou2020dilated}.}}
\end{table*}

\end{document}